\documentclass[10pt]{iopart}
\usepackage{epsfig,graphics,color}

\newcommand \ie {{\it i.e.} }
\newcommand \f {\not\!}

\newcommand \kd  {\delta}
\newcommand \ra  {\rightarrow}

\newcommand \fp {{\bf p}}

\newcommand \fn {{\bf n}}

\newcommand \fx {{\bf x}}

\newcommand \si {\sigma}

\newcommand \x {\cdot}

\newcommand \A {\alpha}

\newcommand \lc {\langle}
\newcommand \rc {\rangle}
\newcommand \prt {\partial}

\newcommand \nt {\noindent}

\newcommand \al {\alpha}

\newcommand \bvec{\left( \begin{array}{c} }
\newcommand \evec{\end{array} \right)}
\newcommand \trc {\mbox{{\bf Tr}}}
  
\newcommand \bea{\begin{eqnarray} }
\newcommand \eea{\end{eqnarray} }
\newcommand \nn {\nonumber}
\newcommand {\be} {\begin{equation}}
\newcommand {\ee} {\end{equation}}
\newcommand {\epem} {$e^+ e^-$}

\newcommand {\gev} {\mbox{GeV}}

\begin{document}

\article[High Pt hadron-hadron correlations]{Hot Quarks 2004}
{High Pt hadron-hadron correlations} 
\author{A. Majumder and Xin-Nian Wang}
\address{Nuclear Science Division, 
Lawrence Berkeley National Laboratory\\
1 Cyclotron road, MS:70R0319, Berkeley, CA 94720}
\begin{abstract} 
We propose the formulation of a dihadron fragmentation function in terms
of parton matrix elements. Under the collinear factorization approximation
and facilitated by the cut-vertex technique, the two hadron inclusive
cross section at leading order (LO) in e+ e- annihilation is shown to
factorize into a short distance parton cross section and the long distance
dihadron fragmentation function. We also derive the DGLAP evolution
equation of this function at leading log. The evolution equation for the
non-singlet and singlet quark fragmentation function and the gluon
fragmentation function are solved numerically with the initial condition
taken from event generators. Modifications to the dihadron fragmentation
function from higher twist corrections in DIS off nuclei are computed.
Results are presented for cases of physical interest.
\end{abstract}



\section{Introduction}


One of the most promising signatures for the formation of a
quark gluon plasma (QGP) in a heavy-ion collision 
has been that of jet quenching \cite{quenching}. This phenomenon 
leads to the suppression 
of high $p_T$ particles emanating from such collisions. Such jet 
quenching phenomena have been among the most striking experimental 
discoveries from the Relativistic Heavy Ion Collider (RHIC) at 
Brookhaven National Laboratory. Simultaneously, the modification 
of jets in cold nuclear matter has been measured via the 
deep-inelastic scattering (DIS) of leptons off nuclei at the 
HERMES detector at the Deutsche Elektronen-Synchrotron (DESY).
Such experiments, besides providing an essential baseline for 
jet modifications in hot matter, are interesting tests of 
nuclear enhanced higher twist effects on parton propagation in 
nuclei.  

In the investigation of jet suppression in heavy-ion collisions, 
correlations between two high $p_T$ 
hadrons in azimuthal angle are used to study the change of jet 
structure~\cite{adl03}. 
While the back-to-back correlations are suppressed in central $Au+Au$ 
collisions, indicating parton energy loss, the same-side 
correlations remain approximately the same as in $p+p$ and $d+Au$ collisions.
An almost identical situation has been observed in the modification 
of fragmentation functions in cold nuclear matter: while the single 
inclusive measurements indicate a large suppression hadrons with large 
momentum fractions (large $z$), 
the ratio of the double inclusive measurements to the 
single inclusive measurements show minimal variation with atomic 
number. 
Given the experimental kinematics, this is considered as an indication
of parton hadronization outside the medium. However, since the
same-side correlation corresponds to two-hadron distribution
within a single jet, the observed phenomenon is highly nontrivial.
To answer the question as to why 
a parton with a reduced energy would give the same two-hadron
distribution, one has to take a closer
look at the single and double hadron fragmentation functions and
their modification in a medium. In this article, we present 
results for the evolution of the dihadron fragmentation function in the
process of $e^+e^-$ annihilation, and its modification in a cold 
nuclear medium. Modifications in a hot QGP will be presented in a 
later effort.

For reactions at an energy scale much above $\Lambda_{QCD}$, 
one can factorize the cross section into a short-distance 
parton cross section which is computable order by order 
as a series in $\al_s(Q^2)$ and a long-distance phenomenological 
object (the fragmentation function)
which contains the non-perturbative information of parton 
hadronization~\cite{col89}. 
These functions are process independent. 
If measured at one energy scale, they may be predicted for all other 
energy scales via the Dokshitzer-Gribov-Lipatov-Altarelli-Parisi (DGLAP) 
evolution equations \cite{gri72}.
In this article, we will extend this formalism to the 
double inclusive fragmentation function $D_q^{h_1,h_2}(z_1,z_2,Q^2)$
or the dihadron fragmentation function, where $z_1,z_2$ represent 
the forward momentum fractions of the two hadrons. In the absence of 
experimental data, the dihadron 
fragmentation functions will be sampled at one energy scale from event 
generators 
and the prediction of the derived DGLAP evolution will be verified against 
the measurement at another energy scale. The dihadron fragmentation 
function will then be subjected to medium modification in a DIS scenario.
Comparisons with data from the HERMES experiment will be presented. 
 

\section{The double fragmentation function and its evolution }


We begin our analysis with the following semi-inclusive process 
\bea
e^+ + e^- \ra \gamma^* \ra h_1 + h_2 + X \nonumber
\eea
of $e^+ e^-$ annihilation.
We consider two-jet events where both the 
identified hadrons $h_1$ and $h_2$ emanate from the same jet. 
At leading order in the strong coupling, this occurs 
from the conversion of the virtual photon into a 
back-to-back quark and anti-quark pair which 
fragment into two jets of hadrons. 

In the limit of very large $Q^2$ of the reaction, we may 
invoke the collinear approximation. Under this approximation, 
at leading twist,  
we can demonstrate the factorization of 
the two-hadron inclusive cross
section into a hard total partonic cross section $\si_0$ and 
the double inclusive fragmentation function $D^{h_1 h_2}_{q}(z_1,z_2)$ (see 
Ref. \cite{maj04} for details),

\bea
\frac{d^2 \si}{dz_1 dz_2} = \sum_{q} \si_0^{q\bar{q}} 
\left[ D_{q}^{h_1 h_2} (z_1,z_2) + 
D_{\bar{q}}^{h_1 h_2} (z_1,z_2) 
\right].
\label{LO_Dz1z2}
\eea

\nt In the above equation, the leading order 
double inclusive fragmentation function of a quark is obtained as

\bea
\!\!\!\!\!\!\!\!\!\!\!\!\!\!D_q^{h_1,h_2}(z_1,z_2) &=& \int 
\frac{dq_\perp^2}{8(2\pi)^2}  \frac{z^4}{4z_1z_2}
\int \frac{d^4 p}{(2\pi)^4}   
\trc \bigg[ \frac{ \f n}{2 \fn \x \fp_h} 
\int d^4 x e^{i\fp \x \fx} \sum_{S - 2} \nn \\
& & 
\lc 0 | \psi_q^\alpha (x) | p_1 p_2 S-2 \rc  
\lc p_1 p_2 S-2 | \bar{\psi}_q^\beta (0) | 0 \rc \bigg]
\kd \left( z - \frac{p_h^+}{p^+}  \right) .\label{dihad_def} 
\eea

\begin{figure}[htb!]
\hspace{0cm}
  \resizebox{2.4in}{2.4in}{\includegraphics[0in,0in][6.0in,6in]{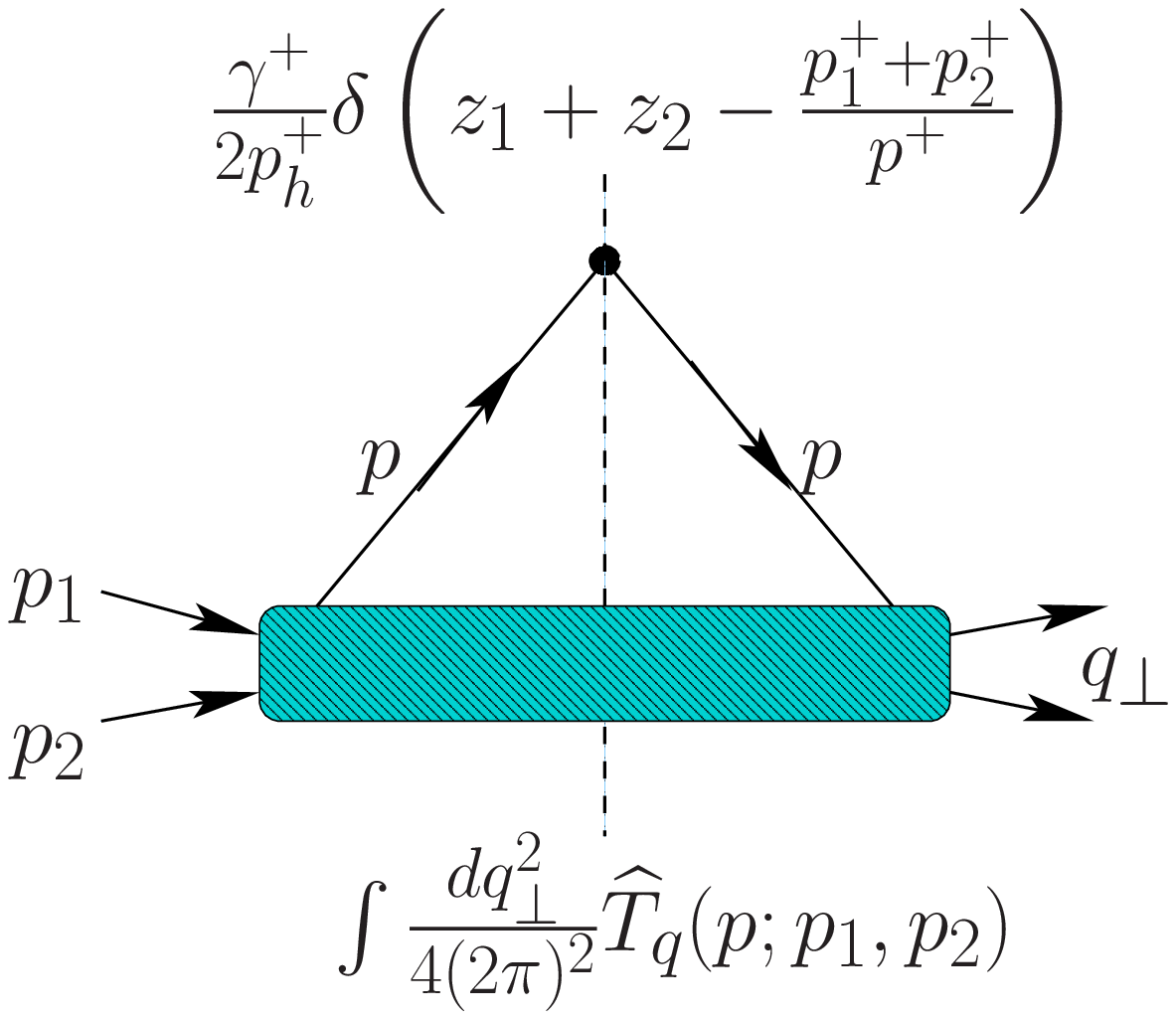}} 
\hspace{0.25cm}
 \resizebox{2.15in}{3.15in}{\includegraphics[0.5in,1in][5.5in,9in]{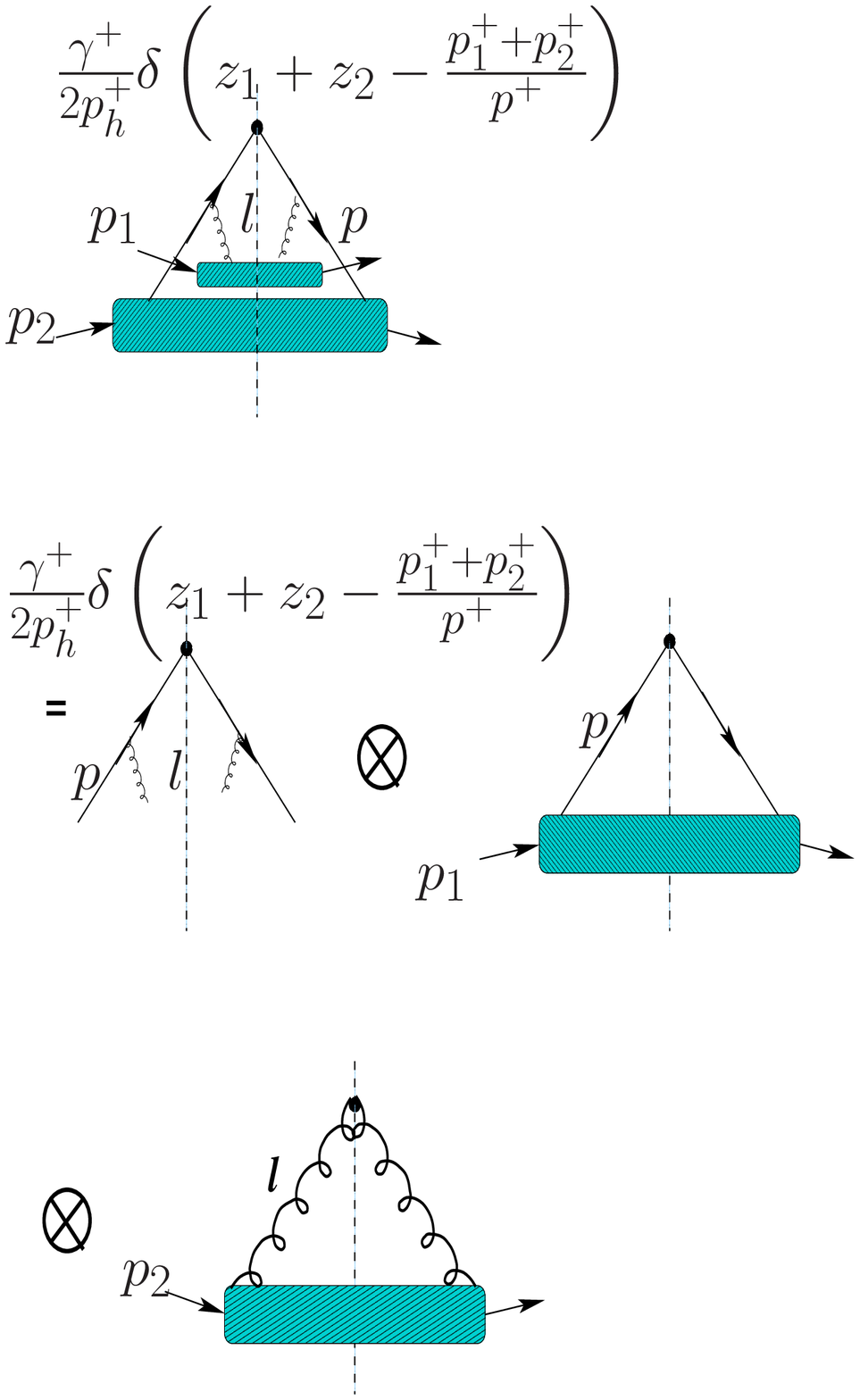}} 
  \caption{The left panel represents the 
  cut-vertex representation of the dihadron fragmentation
    function. The right panel represents a Next-to-Leading order 
    correction from quark and gluon single fragmentation.}
    \label{cutvert1}
\end{figure}

\nt
In the above equation $z=z_1+z_2$, $\fp$ represents the momentum of the 
fragmenting parton, $\fp_h$ is the sum of the momenta of the two 
detected hadrons 
\ie $\fp_1 + \fp_2$, $q_\perp$ is the relative transverse momentum between the 
two detected hadrons and $\fn$ is a light-like null vector. The sum over 
$S$ indicates a sum over all possible final hadronic states.
The above equation 
may be represented by the diagrams of the cut vertex notation \cite{mue78} as 
that in the left panel of Fig.~\ref{cutvert1}. Note that all transverse 
momentum $q_\perp$ up to a scale $\mu_\perp$ 
have been integrated over into the definition of the 
fragmentation function.  Hadrons with transverse momenta $\geq \mu_\perp$
may not emanate from the fragmentation of a single parton.

In the interest of simplicity 
we first specialize to the case of the non-singlet fragmentation function, 
$D_{NS}^{h_1,h_2} (z_1,z_2) =  D_q^{h_1,h_2} (z_1,z_2) -  
D_{\bar{q}}^{h_1,h_2} (z_1,z_2)$.
With the definition of the dihadron fragmentation functions 
in the operator formalism,
which is shown to factorize from the hard parton cross section at LO, 
we may now evaluate its DGLAP evolution 
by computing the double inclusive cross 
section at next to leading order (NLO).
This constitutes a rather involved procedure; the reader is referred to 
Ref.~\cite{maj04} for details. The resulting DGLAP evolution consists of 
two parts. There remains the usual evolution 
due to the radiation of soft gluons and both detected 
hadrons emanating with a small 
transverse momentum off the same parton. Hadron pairs with 
transverse momentum larger than $\mu_\perp$ 
are produced perturbatively, by the independent 
single fragmentation of a quark and a gluon which emanate from the 
splitting of a quark. The contribution of this process to the evolution
of the dihadron fragmentation function is represented by the cut-vertex 
diagram in the right panel of Fig. \ref{cutvert1}. Incorporating this 
diagram, we obtain the DGLAP evolution of the dihadron fragmentation 
function as 

\bea 
\fl \frac{\prt D_{NS}^{h_1 h_2} (Q^2)}{\prt \log{Q^2}} = 
\frac{\A_s}{2\pi} \Bigg[ P_{q\ra q g} * D_{NS}^{h_1 h_2} (Q^2)  
+ \hat{P}_{q \ra q g} 
\bar{*} D_{NS}^{h_1} (Q^2) D_g^{h_2} (Q^2)  + 1 \ra 2 \Bigg], \label{ns_dglap}
\eea

\begin{figure}[htb!]
\hspace{0cm}
  \resizebox{2.5in}{2.5in}{\includegraphics[0.5in,1in][7.5in,9.5in]{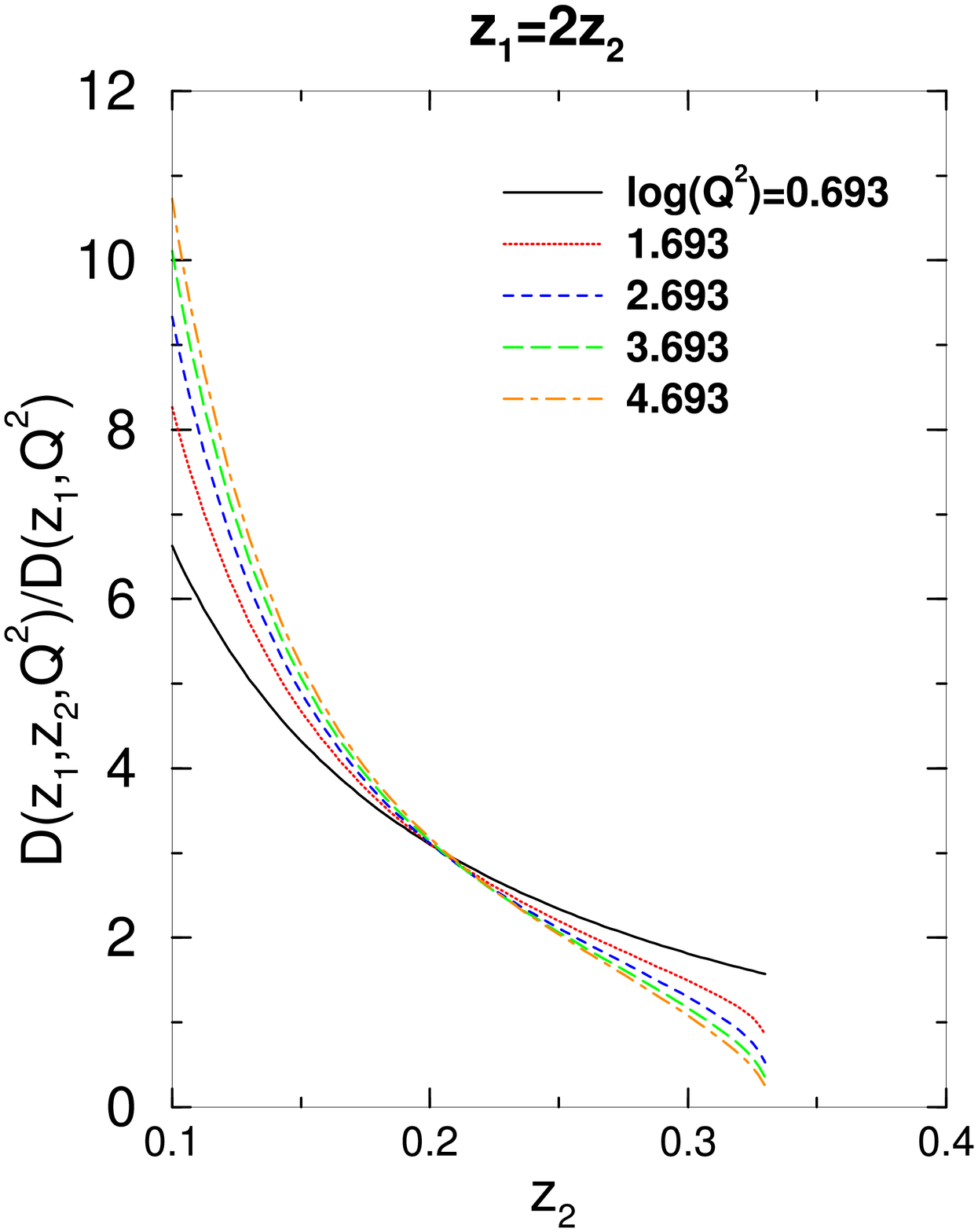}} 
\hspace{0.25cm}
 \resizebox{2.5in}{2.5in}{\includegraphics[0.5in,1in][7.5in,9.5in]{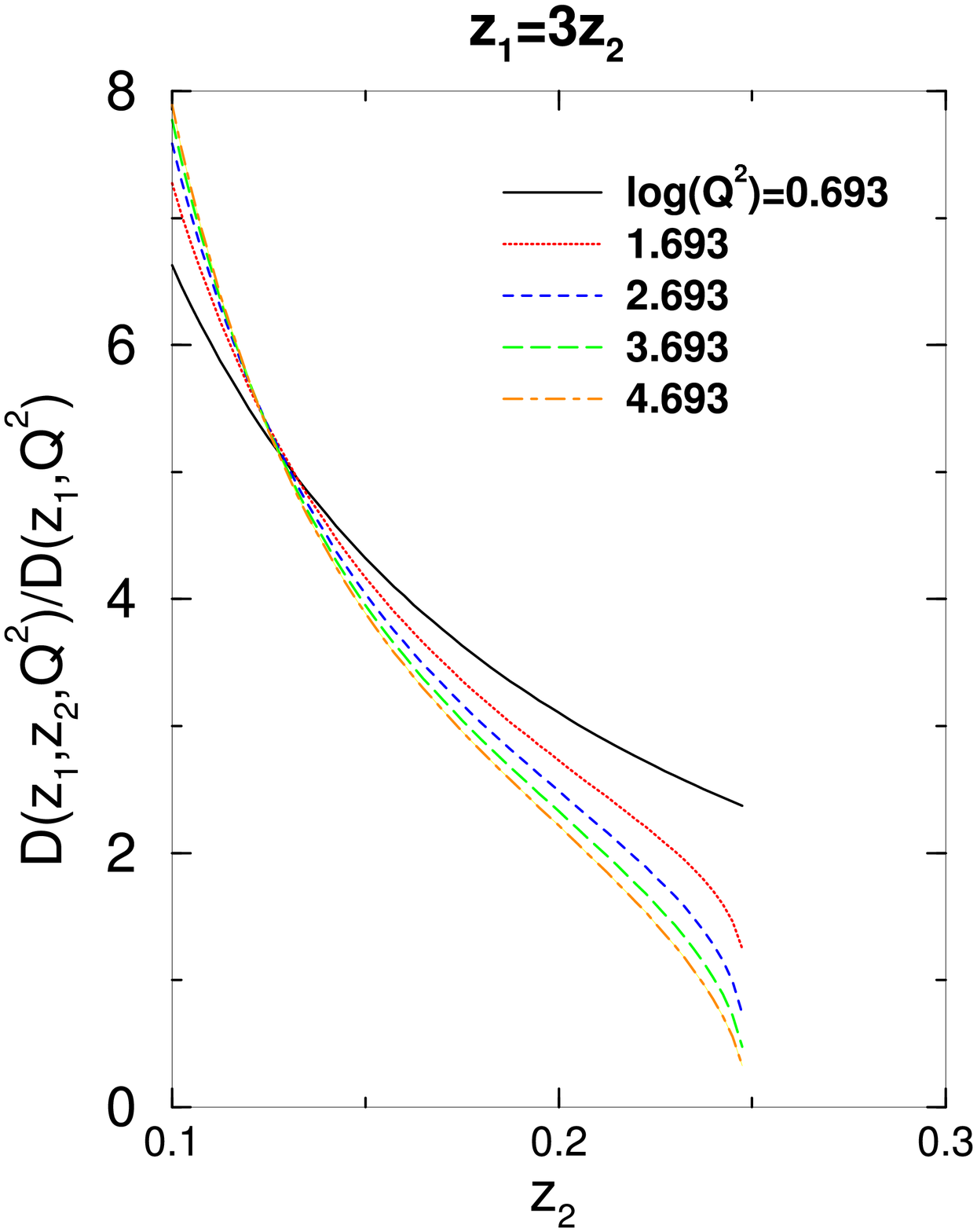}} 
    \caption{Results of the ratio of the non-singlet quark dihadron 
    fragmentation function $D_q^{h_1h_2}(z_1,z_2,Q^2)$ to the single 
    leading fragmentation function  $D_q^{h_1}(z_1,Q^2)$. In the 
    left panel $z_1=2z_2$, in the right panel $z_1=3z_2$. Results are 
    presented from 
    $Q^2= 2 \gev^2 $ to $109 \gev^2$.}
    \label{res3}
\end{figure}

\nt
where the switch $1 \ra 2$ is meant solely for the last term. The 
expressions for the splitting functions ( $P_{q\ra qg}$ and  
$\hat{P}_{q\ra qg}$ ), single fragmentation functions, 
as well as the convolution notations ($*$ and $\bar{*}$)
may be obtained from Ref. \cite{maj04}. The above equation may be solved 
numerically and results for the cases of $z_1=2z_2$ and $3z_2$ are presented in 
Fig.~\ref{res3}. We assume a simple ansatz for the initial condition at 
$Q^2 = 2$ GeV$^2$ \ie $D_{NS}^{h_1 h_2}(z_1,z_2) = D^{h_1}(z_1)D^{h_2}(z_2)$. 
We present results for the evolution with $\log(Q^2)$ at intervals of 1, up to 
$\log(Q^2) = 4.693$ \ie $Q^2 = 109$ GeV$^2$. It may be noted that the results 
demonstrate minimal change as the energy scale is varied over a wide range of 
values.

In generalizing from the non-singlet quark fragmentation function to the 
quark or antiquark fragmentation function one simply includes a 
piece in which the fragmenting quark may radiate a gluon and both 
detected hadrons may emanate from the fragmentation of this gluon.
However, this new piece couples the evolution of the quark fragmentation
function with the evolution of the gluon fragmentation function. The 
evolution of the gluon fragmenation function has the usual components 
where the gluon may radiate further gluons prior to undergoing fragmentation, 
or convert into a quark-antiquark pair with the detected hadrons 
materializing from either parton. It also contains the new additional 
pieces where each of the detected hadrons may emanate from each of the 
radiated partons. We illustrate the various components of the 
gluon fragmentation function by the following equation (see Ref.~\cite{maj04b} 
for details),

\bea 
\fl \frac{\prt D_{g}^{h_1 h_2} (Q^2)}{\prt \log{Q^2}} \!\!\!\!\!\!\!\!&=& 
\frac{\A_s}{2\pi} \Bigg[ \sum_{f = 1}^6 \hat{P}_{g \ra f \bar{f}} 
* D_{f}^{h_1 h_2} (Q^2)  
+  \sum_{f = 1}^3 P_{q \ra f \bar{f} } \bar{*} D_f^{h_1} (Q^2) 
D_{\bar{f}}^{h_2} (Q^2)  + 1 \ra 2  \nn \\ 
\mbox{}\!\!\!\!\!\!\!\!&+& P_{g\ra g g} * D_{g}^{h_1 h_2} (Q^2)  + 
P_{g\ra g g} \bar{*} D_{g}^{h_1} (Q^2) D_{g}^{h_2} (Q^2)  
\Bigg].   \label{g_dglap}
\eea
  
In the above equation, $f$ indicates the flavour of a fermion. In the 
first term, the 
sum runs from 1 to 6 indicating a sum over both quarks and antiquarks.
In the second term the sum runs from 1 to 3 indicating only fermions.
The case of quarks replaced with antiquarks is indicated with the 
switch 1 to 2 which only applies to this term. The remaining terms 
indicate splitting into two gluons followed by fragmentation function.
Expressions for the splitting functions may be obtained from Ref.~\cite{maj04b}. 

\begin{figure}[htb!]
\hspace{0cm}
  \resizebox{2.5in}{2.5in}{\includegraphics[0.5in,1in][7.5in,9.5in]{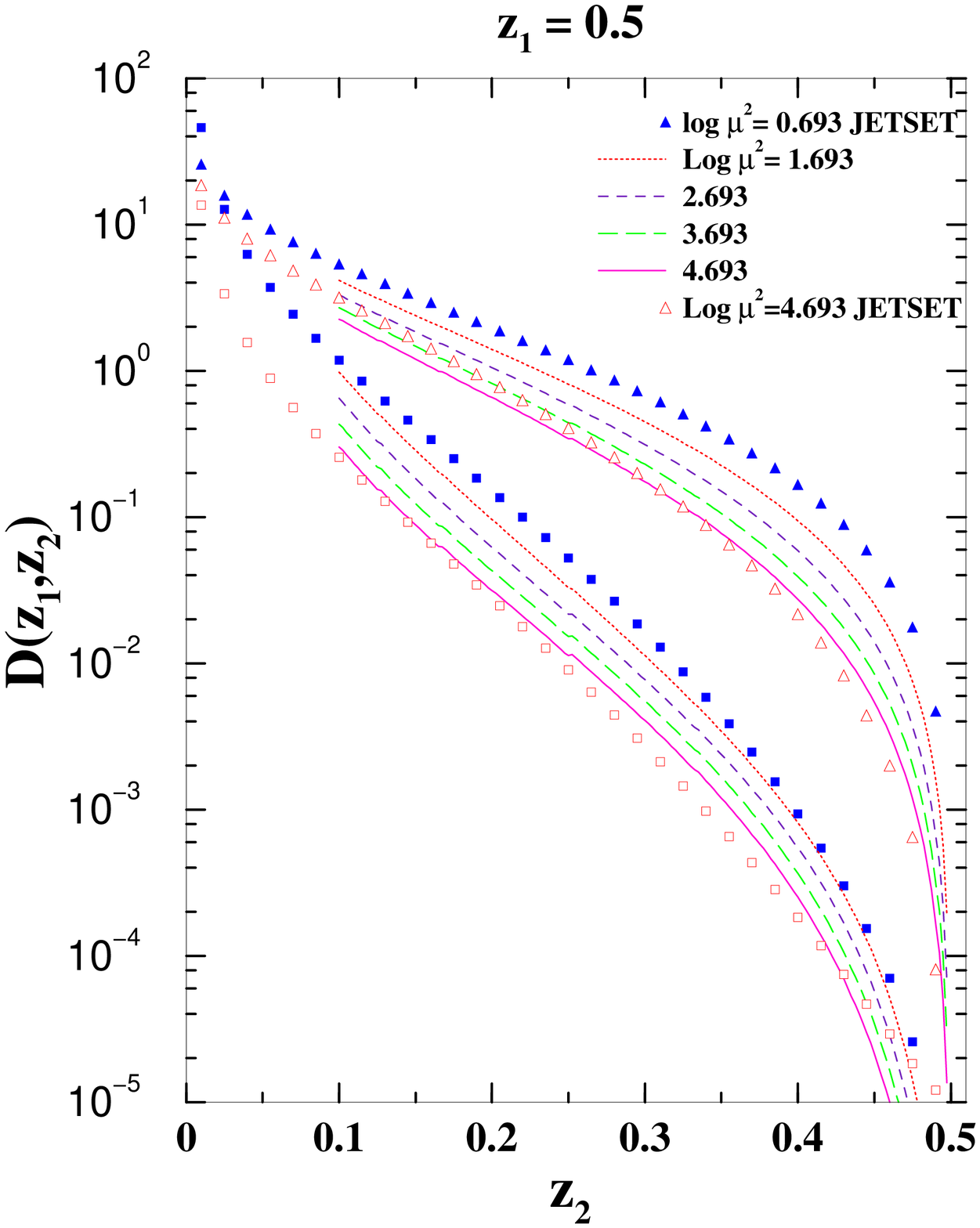}} 
\hspace{0.25cm}
 \resizebox{2.5in}{2.5in}{\includegraphics[0.5in,1in][7.5in,9.5in]{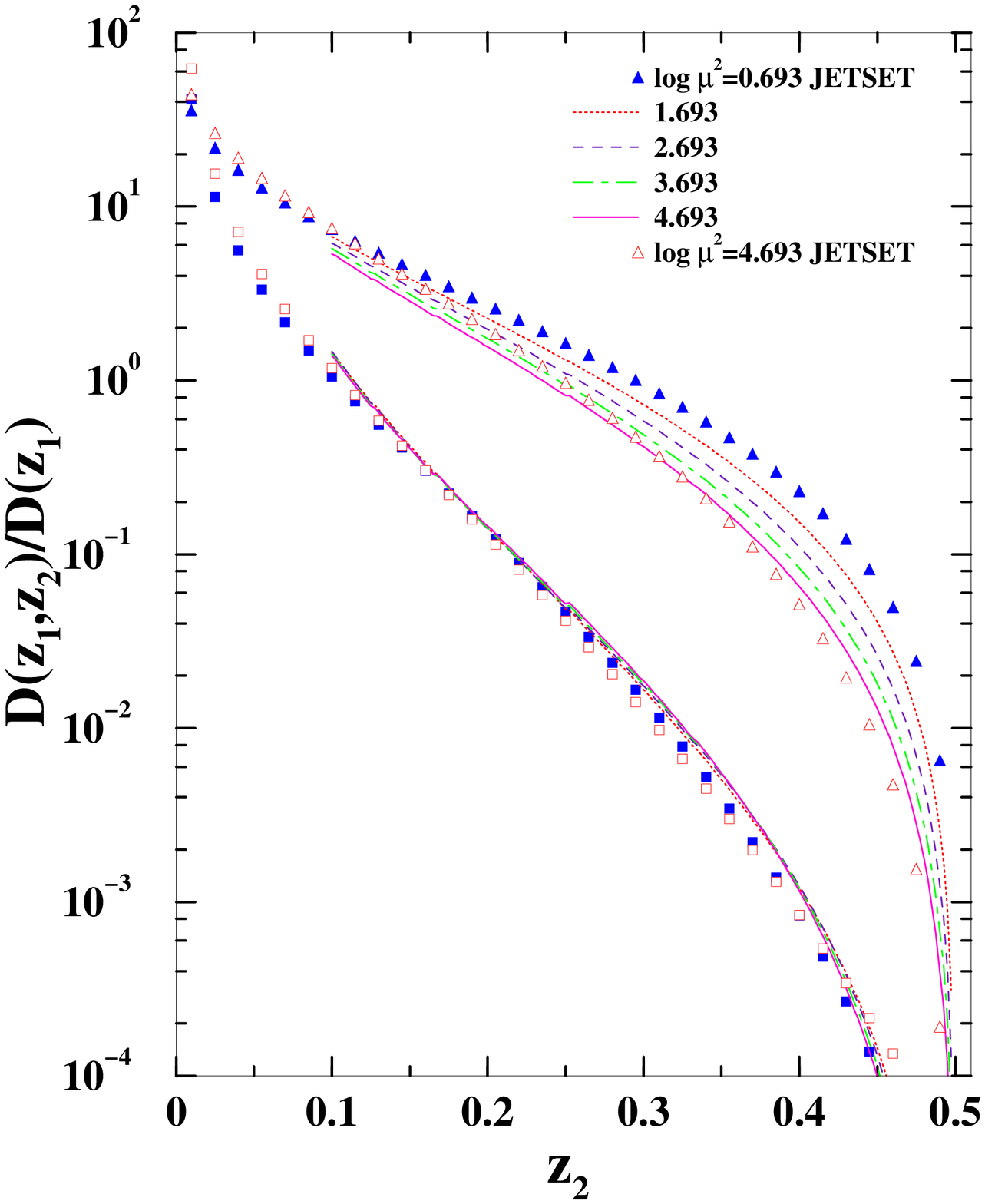}} 
    \caption{Left panel shows results of the evolution of the quark and 
gluon dihadron fragmentation 
functions. 
The right panel is a plot of the ratio of $D^{h_1h_2}(z_1,z_2,Q^2)$ 
to the single leading fragmentation function  $D^{h_1}(z_1,Q^2)$, for the 
cases shown in the left panel. See text for details}
    \label{fig3}
\end{figure}

Once the fragmentation functions are measured at a given 
energy scale, their variation with energy scale can be predicted 
with the aid of the evolution equations mentioned above. To date, 
there remain no measurements of two particle correlations
in \epem collisions. However, Monte Carlo event generators such as 
JETSET have enjoyed great success as simulators of such collisions. 
The quark and gluon dihadron fragmentation at a scale $\mu^2 = 2$GeV$^2$ 
are extracted from two and three jet events in JETSET (see Ref.~\cite{maj04b} 
for details). These are represented by the filled symbols (triangles for quarks and 
squares for gluons) in Fig.~\ref{fig3}. In both cases, the momentum fraction of the
leading hadron is set as $z_1 = 0.5$.
Once measured, these fragmentation functions are then 
evolved with the aid of the evolution equations outlined above. 
Results of the evolution with $\log(Q^2)$ at intervals of 1, up to 
$\log(Q^2) = 4.693$ \ie $Q^2 = 109$ GeV$^2$ are presented in the left panel 
of Fig.~\ref{fig3} as the 
various lines. The fragmentation functions are once again measured at 
the higher scale from JETSET and compared with the calculation (open triangles 
for quarks and open squares for the gluons).  
The excellent agreement between the evolution equation and 
the predictions from JETSET at the higher scale provide an important 
test of the derived DGLAP equation for the dihadron fragmentation function.

The right panel of Fig.~\ref{fig3} shows the variation of the ratio
$D(z_1,z_2,Q^2)/D(z_1,Q^2)$ with $Q^2$ while $z_1$ is held fixed at 
0.5. While the quark dihadron fragmentation function displays a slow 
variation, the gluon fragmentation function shows almost no change 
as the energy scale is varied over two orders of magnitude. This 
is consistent with the results obtained for the evolution of the 
non-singlet functions.

\section{Medium Modification}

The HERMES detector at DESY measures the semi-inclusive cross section 
for the production of one or two hadrons from electroproduction 
off nuclei such as Nitrogen and Krypton \cite{din04}. 
The measured quantity, $R_A^{h_1 h_2}$, is the number of events with at least two 
detected hadrons with the leading hadron assuming a momentum fraction  
of $z_1 \geq 0.5$, divided by the number of events with at least one 
hadron with a momentum fraction $z \geq 0.5$. 


Due to various detector systematics, a double ratio is 
presented: $ R_{A/D}=R_A^{h_1 h_2}/R_D^{h_1 h_2},$
where $R_D^{h_1 h_2}$ is the same ratio as discussed above but measured in 
scattering off Deuterium. Assuming that the Deuterium nucleus is two 
small to induce any noticeable medium effects on the fragmentation of 
hard partons passing through it and in the limit of low pair multiplicity 
per event, 
we equate the above ratio as 

\bea
R_{A/D} = \frac{\tilde{D}_q^{h_1 h_2}(Q^2) / \tilde{D}_q^{h_1}(Q^2) }
{D_q^{h_1 h_2}(Q^2) / D_q^{h_1}(Q^2) },
\eea 

\begin{figure}[htb!]
\hspace{0cm}
  \resizebox{2.5in}{2.5in}{\includegraphics[1.0in,6.5in][4.75in,10in]{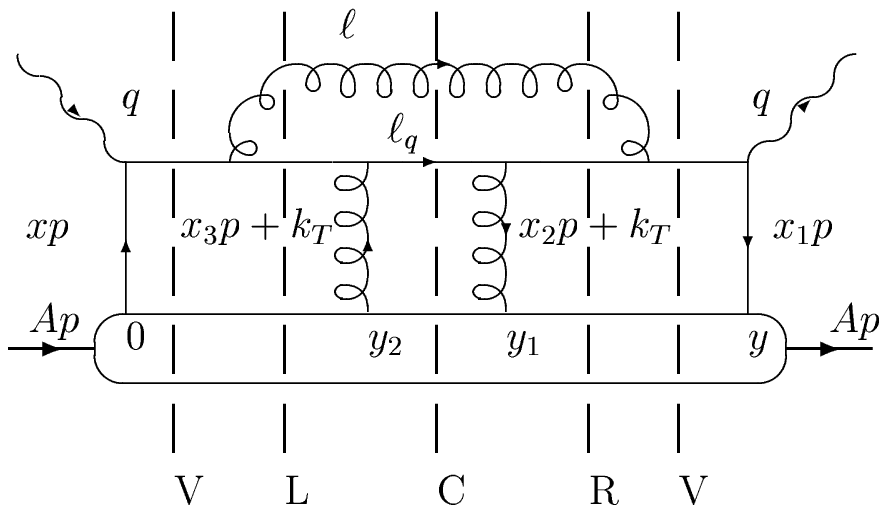}} 
\hspace{0.25cm}
 \resizebox{2.5in}{2.5in}{\includegraphics[0.5in,1in][7.5in,10in]{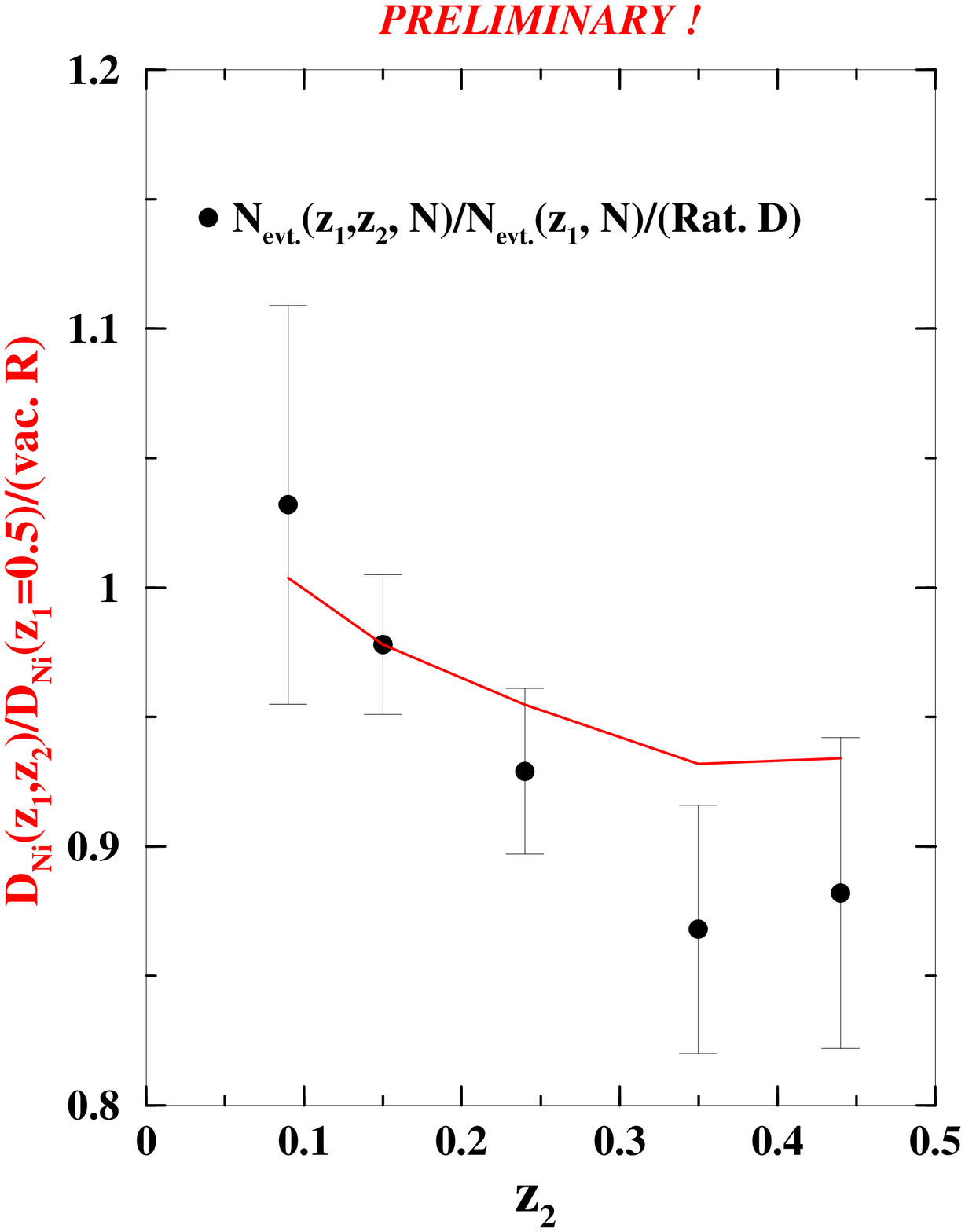}} 
  \caption{Left panel shows one of the higher twist diagrams involved in the 
calculation of the medium modification of the fragmentation functions. The various 
dashed lines indicate the various cuts: left (L), central (C), right (R) and virtual (V).
The right panel shows the experimental data from HERMES for $R_{A/D}$ and the solid 
line is the calculation. Both the data and the calculation are preliminary.}
    \label{fig4}
\end{figure}

\nt
where $D_q^{h_1 h_2}, D_q^{h_1}$  represent the quark fragmentation 
functions in vacuum, and $\tilde{D}_q^{h_1 h_2}, \tilde{D}_q^{h_1}$ represent
those modified by the medium. Values of $R_{A/D}$ for Nitrogen are 
plotted in the right panel of Fig.~\ref{fig4}.

The calculation of the medium modification of the fragmentation 
function is carried out in the nuclear enhanced higher twist 
formalism. This formalism is presented in detail in Ref.~\cite{guowang}.
In principle, one calculates the cross section for the process 
where the recoiling lepton imparts a large momentum transfer ($Q^2$) to 
one of the quarks in the incoming nucleus via single photon exchange. The 
struck quark then undergoes multiple scattering off the soft gluon 
field of the nucleus prior to exiting the medium and fragmenting to hadrons. 
In this process it absorbs a certain amount of 
transverse momentum and virtuality, which it loses by gluon bremsstrahlung.
Alternatively, the struck quark may be produced off shell and may 
radiate gluons prior to scattering off the soft gluons. In either 
case the radiated gluon (which also carries a colour charge) may 
also encounter further scattering of the gluon field prior to 
exiting the medium and fragmenting into a jet of hadrons. Interferences 
between the various processes mentioned above also need to be considered.

The eventual calculation of the medium modification involves the 
computation of multiple higher twist diagrams such as the one displayed in the
left panel of Fig.~\ref{fig4}. While these diagrams, which involve quark gluon 
correlations in nuclei, are suppressed 
by powers of $Q^2$, they receive an enhancement $\sim A^{\frac{1}{3}}$ 
from the fact that the quark and gluon may originate from any of the $A$
nucleons in the nucleus (for further details see Refs.~\cite{guowang,maj04c}).
Calculations for a Nitrogen nucleus are presented and compared to the 
HERMES data in the right panel of Fig.~\ref{fig4}.

\section{Discussions and Conclusions}

This study was motivated by the observation \cite{adl03} that the
same side correlations of two high $p_T$ hadrons in central
$Au+Au$ collisions remain approximately unchanged as compared
with that in $p+p$ collisions. A similar situation was also observed 
in DIS experiments. 
Neglecting the differences
in production cross section and fragmentation functions for
different parton species, the integrated yield of the same side 
correlation over a small range of angles should be the ratio of 
the dihadron to the single hadron fragmentation functions, 
$D_a^{h_1h_2}(z_1,z_2,Q^2) / D_a^{h_1}(z_1,Q^2)$.
Thus, we evaluated the influence of DGLAP evolution on the 
ratio $D_q^{h_1h_2}(z_1,z_2,Q^2) / D_q^{h_1}(z_1,Q^2)$. 
Our numerical results indeed show little change of the ratio 
as $Q^2$ is varied over a wide range of values for both singlet and 
non-singlet quarks and for gluons.

We evaluated the influence of medium modification due to 
multiple scattering and induced gluon radiation on the 
above ratio in a cold nuclear medium. The results were 
found to be consistent, both with the experiment and also the 
vacuum evolution of the ratio. Calculations for the 
modification in a deconfined medium are in progress.

\end{document}